\documentclass[sigplan,screen]{acmart}
\usepackage{microtype}
\usepackage{arydshln}
\usepackage{amsmath,amsfonts}
\usepackage{algorithmic}
\usepackage{graphicx}
\usepackage{textcomp}
\usepackage{xcolor}
\usepackage{xspace}
\usepackage{enumitem}
\newcommand{\tool}{\emph{FGVulDet}\xspace}
\usepackage{amsmath}
\usepackage{listings}
\usepackage{multicol}
\usepackage{color} 
\usepackage{caption}
\usepackage{multirow}
\usepackage[linesnumbered,ruled,vlined]{algorithm2e}
\usepackage{subcaption}
\usepackage[most]{tcolorbox}
\usepackage{hyperref}
\definecolor{codegreen}{rgb}{0,0.6,0}
\lstdefinelanguage{diff}{
  morecomment=[f][\color{blue}]{@@},  morecomment=[f][\color{red}]-, 
  morecomment=[f][\color{codegreen}]+,
  morecomment=[f][\color{red}]{---}, 
  morecomment=[f][\color{codegreen}]{+++},
}
\setcopyright{rightsretained}
\acmDOI{10.1145/3652032.3657564}
\acmYear{2024}
\copyrightyear{2024}
\acmSubmissionID{pldiws24lctesmain-p2-p}
\acmISBN{979-8-4007-0616-5/24/06}
\acmConference[LCTES '24]{Proceedings of the 25th ACM SIGPLAN/SIGBED International Conference on Languages, Compilers, and Tools for Embedded Systems}{June 24, 2024}{Copenhagen, Denmark}
\acmBooktitle{Proceedings of the 25th ACM SIGPLAN/SIGBED International Conference on Languages, Compilers, and Tools for Embedded Systems (LCTES '24), June 24, 2024, Copenhagen, Denmark}
\received{2024-02-29}
\received[accepted]{2024-04-01}

\begin{document}
\title{Enhancing Code Vulnerability Detection via Vulnerability-Preserving Data Augmentation}

\author{Shangqing Liu}
\orcid{0000-0002-5598-4006}
\affiliation{%
  \institution{Nanyang Technological University}
  \country{Singapore}
}
\email{liu.shangqing@ntu.edu.sg}

\author{Wei Ma}
\orcid{0000-0002-0044-466X}
\authornote{Corresponding author}
\affiliation{%
  \institution{Nanyang Technological University}
  \country{Singapore}
}
\email{ma_wei@ntu.edu.sg}

\author{Jian Wang}
\orcid{0000-0002-0393-3709}
\affiliation{%
  \institution{Nanyang Technological University}
  \country{Singapore}
}
\email{jian004@e.ntu.edu.sg}

\author{Xiaofei Xie}
\orcid{0000-0002-1288-6502}
\affiliation{%
  \institution{Singapore Management University}
  \country{Singapore}
}
\email{xfxie@smu.edu.sg}

\author{Ruitao Feng}
\orcid{0000-0001-9080-6865}
\affiliation{%
  \institution{Singapore Management University}
  \country{Singapore}
}
\email{rtfeng@smu.edu.sg}

\author{Yang Liu}
\orcid{0000-0001-7300-9215}
\affiliation{%
  \institution{Nanyang Technological University}
  \country{Singapore}
}
\email{yangliu@ntu.edu.sg}





\begin{abstract}
Source code vulnerability detection aims to identify inherent vulnerabilities to safeguard software systems from potential attacks. Many prior studies overlook diverse vulnerability characteristics, simplifying the problem into a binary (0-1) classification task for example determining whether it is vulnerable or not. This poses a challenge for a single deep-learning based model to effectively learn the wide array of vulnerability characteristics. Furthermore, due to the challenges associated with collecting large-scale vulnerability data, these detectors often overfit limited training datasets, resulting in lower model generalization performance.

To address the aforementioned challenges, in this work, we introduce a fine-grained vulnerability detector namely \tool. Unlike previous approaches, \tool employs multiple classifiers to discern characteristics of various vulnerability types and combines their outputs to identify the specific type of vulnerability. Each classifier is designed to learn type-specific vulnerability semantics. Additionally, to address the scarcity of data for some vulnerability types and enhance data diversity for learning better vulnerability semantics, we propose a novel vulnerability-preserving data augmentation technique to augment the number of vulnerabilities. Taking inspiration from recent advancements in graph neural networks for learning program semantics, we incorporate a Gated Graph Neural Network (GGNN) and extend it to an edge-aware GGNN to capture edge-type information. \tool is trained on a large-scale dataset from GitHub, encompassing five different types of vulnerabilities. Extensive experiments compared with static-analysis-based approaches and learning-based approaches have demonstrated the effectiveness of \tool.
\end{abstract}




\maketitle

\section{Introduction}
Software vulnerability is defined as a weakness in the software system that could be exploited by a threat source. With the increasing number of open-source libraries and the expanding size of software systems, the count of software vulnerabilities has been rising rapidly. Since these vulnerabilities can be exploited by malicious attackers, causing significant financial and social damages, vulnerability detection and patching have garnered widespread attention from academia and industry. For instance,  the Common Vulnerabilities and Exposures (CVE) Program and the National Vulnerability Database (NVD) have been established to identify and patch vulnerabilities before they are exploited. So far, over 100,000 vulnerabilities have been indexed. However, in contrast to the quantity of open-source projects and the speed of software iteration, the number of exploited vulnerabilities is insufficient. In other words, there exists a large number of "silent" vulnerabilities that have not been exploited.

Automated software vulnerability detection remains a crucial yet far from the settled problem. Several techniques have been developed to detect vulnerabilities including static analysis~\cite{du2019leopard, vanegue2013towards, viega2000its4}, fuzzing~\cite{chen2018hawkeye, wang2017skyfire}, symbolic execution~\cite{babic2011statically, stephens2016driller, cha2015program}. Static analysis for vulnerability detection aims to identify vulnerabilities in the source code without execution, typically requiring substantial manual effort from security experts to craft rules. This approach has limited generalization ability across diverse vulnerabilities. Dynamic techniques, such as fuzzing and symbolic execution, identify vulnerabilities by dynamically executing programs. Dynamic approaches demonstrate relatively high precision in vulnerability detection, but configuring execution is complex, and execution results may be incomplete since not every program path can be executed.

Due to the capability of deep learning-based techniques to automatically extract features, more research focuses on utilizing DL techniques for vulnerability detection~\cite{li2018vuldeepecker, zou2019muvuldeepecker, russell2018automated, dam2017automatic, li2021sysevr, lin2019software, zhou2019devign}. In early DL-based vulnerability detection works, some works~\cite{russell2018automated} employed convolutional neural networks (CNNs) to leverage their powerful convolution capabilities for learning vulnerability-related features. However, as programs are not fixed length compared to images, they are not well-suited for CNNs. To avoid this problem, some other works~\cite{dam2017automatic, li2018vuldeepecker, zou2019muvuldeepecker, lin2019software} treat programs as a flat sequence and apply recurrent neural networks (RNNs) with Long Short-Term Memory (LSTM) directly to learn the vulnerability features. Yet, in vulnerability scenarios, certain types of vulnerabilities, such as buffer overflow, are related to data flow, which cannot be captured by the program text alone. To capture the data dependency and control dependency of programs, Li et al.~\cite{li2021sysevr} proposed a program slicing algorithm based on the program dependency graph (PDG) to slice related statements and feed them to Bidirectional RNNs for learning. However, it still fundamentally treats programs as sequences.

\textit{How to learn well-structured control and data dependencies in programs?} Devign~\cite{zhou2019devign} proposed an effective way by encoding programs into a code property graph (CPG) and utilizing this graph through GGNN~\cite{li2015gated} for vulnerability detection, achieving state-of-the-art performance. Afterward, there is a great number of works using GNNs to learn program semantics for source code vulnerability detection~\cite{cheng2019static, wang2020combining, nguyen2022regvd, wang2020combining}. However, most of these works combined various types of vulnerabilities to train a single classifier for vulnerability detection. Moreover, data augmentation has been shown to significantly improve performance on image data~\cite{hendrycks2019augmix, gong2020maxup, he2016deep}. Recent works~\cite{jain2020contrastive, liu2023contrabert} propose to augment code with the same functionality variants by the transformations for contrastive pre-training to learn code functionality for different downstream tasks. However, the defined transformations are at the granularity of the function and it cannot guarantee vulnerability-preserving attribution for vulnerabilities when transforming a function to the variants. Hence, how to perform data augmentation meanwhile preserving the source code vulnerability is a challenge. 

To address these challenges, in this paper, we propose \tool, which is a fine-grained vulnerability detector. Specifically, we train multiple classifiers via the enhanced GGNN for each type of vulnerability on the real collected vulnerability data set from GitHub. Then each model prediction result is ensembled for voting to give the final prediction. Furthermore, we propose a novel vulnerability-preserving data augmentation to enrich the diversity of data and improve the prediction performance. On the side of GGNN, we adopt it and further encode the edge type information along with the node features during message passing i.e., edge-aware GGNN for the enhancement of the vulnerability detection. We claim that the edge type information i.e., ``Flow to'', ``Control'' represents different semantics of programs, and encoding them explicitly during the learning process can facilitate learning more accurate code representations. An extensive evaluation is conducted on five different types of vulnerabilities compared with some static-analysis tools and deep-learning based vulnerability detection approaches have confirmed the superiority of our proposed approach. Further ablation study also reveals the effectiveness of each component in \tool. Our contributions are as follows.
\begin{itemize}[leftmargin=*]
    \item We propose a novel vulnerability-preserving data augmentation technique to enrich the amount of the collected data and mitigate the limitations of rare vulnerabilities in quantity. 
    \item We adopt an edge-aware GGNN by incorporating edge-type features with node features to improve the learning capacity of GGNN for vulnerability detection.
    \item We conduct extensive experiments on the real collected vulnerability data to illustrate the effectiveness of \tool. 
\end{itemize}
\section{Background}\label{sec:background}
\subsection{Problem Definition}
Existing works~\cite{zhou2019devign, duan2019vulsniper, russell2018automated, ndss18vuldeepecker} define source code vulnerability identification as a binary $\{$0,1$\}$ classification problem i.e., labelling all vulnerable functions as 1, regardless of the vulnerable type of the function, which is coarse for vulnerability detection. Differently, in this work, we focus on investigating a fine-grained vulnerability identification problem i.e., for different types of vulnerability, our goal is to learn the corresponding prediction function. Specifically, given a dataset $D=\{D_1, D_2,...,D_t\}$, where $D_t$ is the sub-dataset in D for the vulnerability type $t \in T$ and $T$ is a set of source code vulnerability types, we aim at learning the mapping $f_t \in F $ over $D_t$ to predict whether the function has the vulnerability of type $t$ and $F=\{f_1, f_2,...,f_t\}$ is the prediction function for different vulnerability type. Furthermore, $D_t=\{(c_t, y) | c_t \in \mathcal{C}_t, y \in \mathcal{Y}\}$, where $\mathcal{C}_t$ is a set of functions which contains the vulnerable functions with the vulnerability type $t$ and the corresponding fixed functions and $y = \{0, 1\}$ is the label set with 1 for the vulnerability and 0 for the non-vulnerability.

\subsection{Code Property Graph}
\begin{figure}[t]
    \centering
    \includegraphics[width=0.45\textwidth]{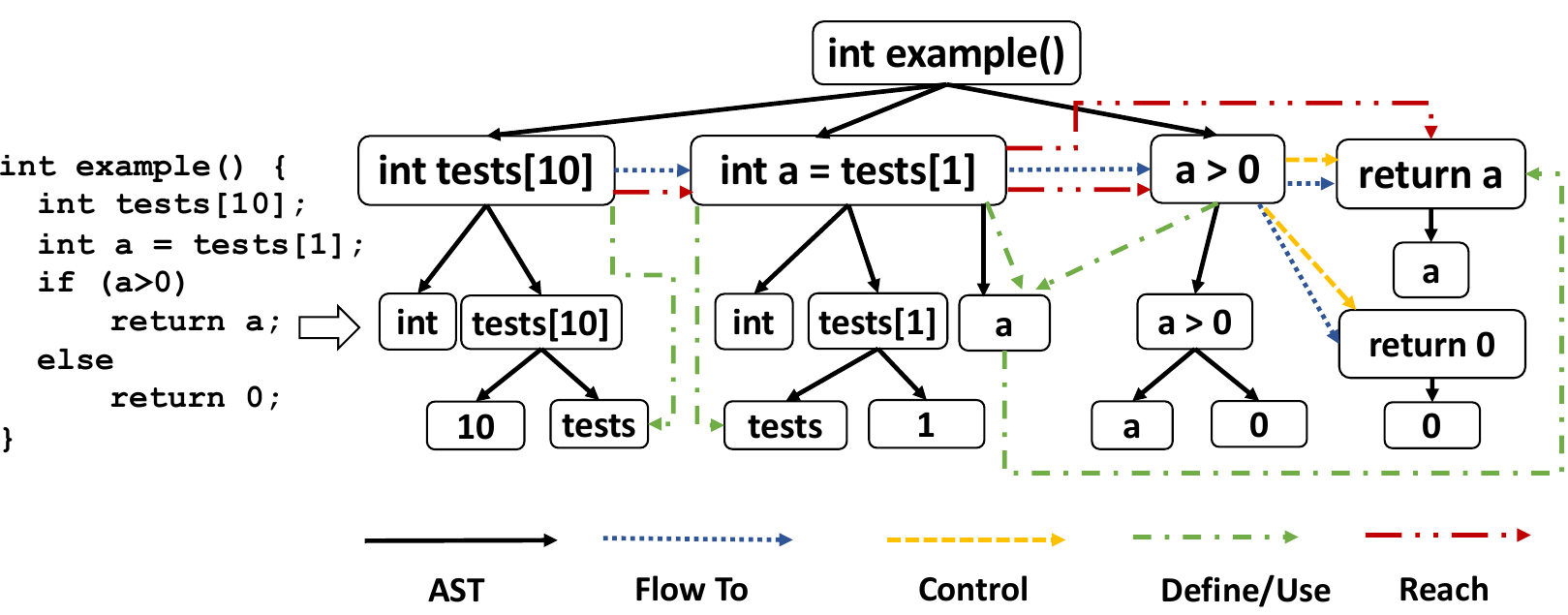}
    \caption{An example to illustrate Code Property Graph.}
    \label{fig:cpg}
\end{figure}

Code property graph (CPG) proposed by Yamaguchi et al.~\cite{yamaguchi2014modeling}, combines several program representations e.g., Abstract Syntax Tree (AST), Control Flow Graph (CFG), Program Dependency Graph (PDG) into a joint graph to represent a program. An illustrated example is shown in Figure~\ref{fig:cpg}. We can observe that AST nodes (defined as black arrows in the graph) are the backstone of CPG. Besides AST, some other semantic representations i.e., control flow, and program dependency information can also be constructed on AST to represent different semantics of the program. For example, CFG represents the statement execution order of the program, and ``Flow To'' (blue arrow) represents this flow order in CPG. Furthermore, PDG is also involved in CPG, and the edges ``Define/Use'' (green arrow), ``Reach'' (red arrow) define the data dependencies, and the ``Control''(yellow arrow) is the control dependency of a program.

\vspace{2mm}


\subsection{Graph Neural Networks}\label{sec:gnn}
Graph Neural Networks (GNNs)~\cite{kipf2016semi, li2015gated} have been widely employed in modeling non-Euclidean data structure such as social networks~\cite{hamilton2017inductive, kipf2016semi}, protein-protein interaction networks~\cite{fortune17}. The primary objective of a GNN is to identify patterns in graph data, relying on information within the nodes and their interconnectedness. There exist various GNN variants, here we describe the broad category of message-passing neural networks~\cite{gilmer2017neural}. 
Suppose the original data can be modelled by a multi-edged graph, denoted as $G=(\mathcal{N},\mathcal{E})$, where $\mathcal{N} = \{n_i\}$ is the node set and 
$\mathcal{E}$ is a set of directed edges $n_i \xrightarrow{k} n_j$ and $k$ is the edge type. Each node $n_i$ is endowed with vector representation $\mathcal{\boldsymbol h}_{n_i}^t$ indexed over a timestep (hop) namely $t$. The node states are updated as

\begin{equation}
\small
    \boldsymbol h_{n_i}^{t+1} = f_t \Bigg(\boldsymbol h_{n_i}^{t}, \bigoplus_{\forall n_j: n_i \xrightarrow{k} n_j} \Big (m^t(\boldsymbol h_{n_i}^{t}, k, \boldsymbol h_{n_j}^{t}) \Big) \Bigg)
    \label{Eq:gnn}
\end{equation}
where $m^t(\cdot)$ is a function that computes the message based on the edge label $k$. $\oplus$ is an aggregation operator that summarises the message from its neighbors and $f_t$ is the update function that updates the state of node $n_i$. The initial state of each node $\boldsymbol h_{n_i}^{0}$ is from node-level information. Equation~\ref{Eq:gnn} updates all node states in a total number of $T$ times recursively and at the end of this iteration, each $\boldsymbol h_{n_i}^{T}$ represents information about the node and how it belongs with the context of the graph. The well-known Graph Convolution Network(GCN)~\cite{kipf2016semi}, Gated Graph Neural Network(GGNN)~\cite{li2015gated} also follow Equation~\ref{Eq:gnn}, but the definitions of $f_t$ and $m^t(\cdot)$ are different. For example, GGNN, which has been widely used in modeling source code~\cite{zhou2019devign, liu2020unified, liu2021retrievalaugmented, learning_to_represent}, employs a single GRU cell~\cite{cho2014learning} for $f_t$, i.e., $f_t=GRU(\cdot, \cdot)$, $\oplus$ is a summation operation and $m^t(\boldsymbol h_{n_i}^{t}, k, \boldsymbol h_{n_j}^{t})=\boldsymbol E_k \boldsymbol h_{n_j}^t$, where $\boldsymbol E_k$ is a learned matrix. The difference between GCN and GGNN lies in $f_t$ is ReLU function~\cite{nair2010rectified} and $\boldsymbol h_{n_i}^{t+1}$ can be expressed as following equation:
\begin{equation}
    \boldsymbol h_{n_i}^{t+1} = \mathrm{ReLU} \Big(\boldsymbol E_t( \boldsymbol h_{n_i}^{t} + \sum_{\forall n_j: n_i \rightarrow n_j} \boldsymbol h_{n_j}^{t})\Big)
\end{equation}


\section{Approach}\label{sec:approach}

\vspace{2mm}
\begin{figure*}[t]
    \centering
    \includegraphics[width=1\textwidth]{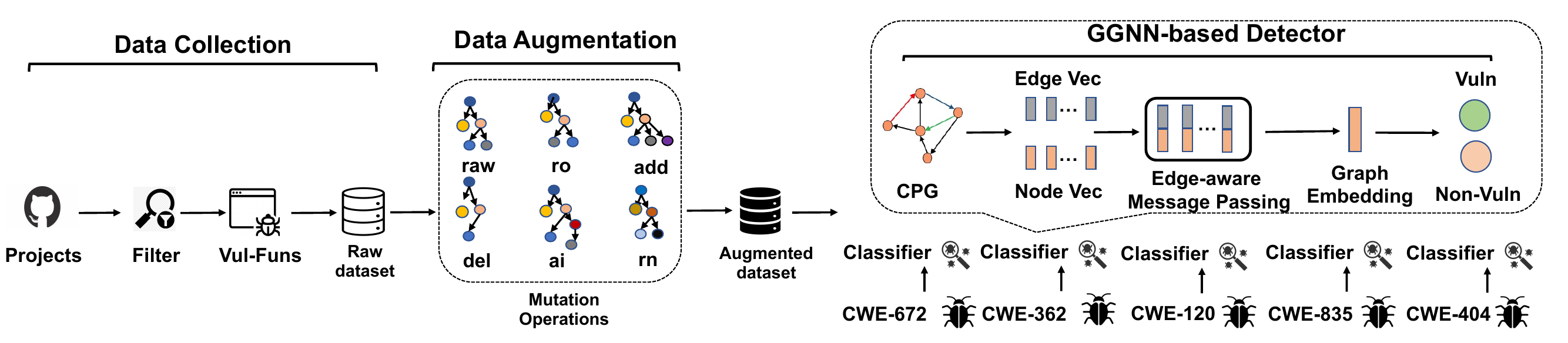}
    \caption{The framework of \tool.}
    \label{fig:framework}
\end{figure*}

\subsection{Overview}
The framework of our approach is illustrated in Figure~\ref{fig:framework}, comprising three main components: Data Collection, which constructs a raw dataset with various types of vulnerabilities from commits; Vulnerability-preserving Data Augmentation, which enhances the original dataset with five mutation operations using a carefully designed vulnerability-preserving slicing algorithm to maintain the original vulnerability semantics and diversify the data; Edge-aware GGNN, which extends the current state-of-the-art GGNN by integrating edge type features into node features during message passing for model learning. During model training, we train multiple binary classifiers for different types of vulnerabilities. In the prediction phase, each classifier provides a prediction result, and \tool aggregates their results through voting to obtain the final prediction.

\subsection{Data Collection}\label{sec2b}
Collecting high-quality datasets of vulnerable functions, especially encompassing various types of vulnerabilities, poses a significant challenge that necessitates expertise. In this work, we propose an effective method for collecting and labeling diverse types of vulnerable and non-vulnerable data. The process involves first gathering commits related to vulnerabilities, followed by extracting pairs of functions from these commits: the vulnerable version $f_v$ and the patched version $f_p$, representing vulnerable and non-vulnerable functions, respectively. The detailed procedures are as below.

\subsubsection{Vulnerability-Related Commit Collection}
To assemble a sizable and diverse dataset of vulnerable functions, we initiate the process by gathering commits from 1614 C-language open-source projects hosted on GitHub. These projects are chosen for their popularity among developers and their diversity in functionality, spanning various domains such as operating systems, networking, and database applications (e.g., Linux Kernel, OpenSSL, QEMU).

To ensure the quality of data labelling, we follow three steps to collect vulnerability-related Commits.
\begin{itemize}[leftmargin=*]
    \item  \textit{Commit Filtering.} We employ vulnerability-related keywords (shown in Table~\ref{keywords}), which have been analyzed and summarized by a team of professional security researchers from a large number of commits. These keywords include five Common Weakness Enumerations (CWE) defined in the National Vulnerability Database (NVD), with each vulnerability type having one or more associated keywords. Commits whose messages do not match any of the keywords in Table~\ref{keywords} are excluded, and the remaining commits are considered more likely to be related to vulnerabilities. For example, in Figure~\ref{fig:patch}, the vulnerability-related commit is accurately captured by the keyword 
    \item \textit{Type Matching.} Commits matched by keywords of multiple vulnerability types are excluded, as we cannot determine which vulnerability type they belong to. We retain commits matched by a single vulnerability type and use that type to label the commits.
    \item \textit{Commit Pruning.} There are some vulnerability-related commits that may modify multiple functions, and not all of these functions are related to the vulnerability. We cannot automatically identify which function is related to the vulnerability, to alleviate this problem, we exclude those commits that modify more than one function. After the above three steps, we obtain a high-quality commit data set with vulnerability type labels.
\end{itemize}

\vspace{2mm}
\begin{table}[!t]
\centering
\caption{Keywords of Five Vulnerability Types.}
\scriptsize	
\begin{tabular}{lll}
\toprule 
\textbf{CWE}& \textbf{Vulnerability} & \textbf{Keywords} \\ \hline
\textit{CWE-404} &\textit{Memory Leak} & 
\begin{tabular}[c]{@{}l@{}}memory leak, information leak, info leak, \\leak info, memory disclosure, leak memory, \\ leak information \end{tabular} \\ \hline
\textit{CWE-835} &\textit{Infinite Loop} & 
 \begin{tabular}[c]{@{}l@{}} infinite loop, endless loop, long loop,\\infinite recursion, deep recursion \end{tabular} \\ \hline
\textit{CWE-120} &\textit{Buffer Overflow} &
\begin{tabular}[c]{@{}l@{}} buffer overflow \end{tabular}  \\ \hline
\textit{CWE-672} &\textit{Operation After Free} &
 \begin{tabular}[c]{@{}l@{}}double free, double-free, DF, use after free,\\ use-after-free, UAF \end{tabular} \\ \hline
\textit{CWE-362} &\textit{Race Conditions} & 
 \begin{tabular}[c]{@{}l@{}}race conditions\end{tabular} \\ \bottomrule 
\end{tabular}
\label{keywords}
\end{table}

\subsubsection{Vulnerable/Non-vulnerable Function Extraction}\label{sec3.2.2}
Given the vulnerability-related commit as input, we can get its corresponding security patch $P_{v}$. We extract vulnerable functions $f_v$ and patched functions $f_p$ based on the change statements (i.e., the added statements $S_{add}$ and the deleted statements $S_{del}$) from $P_{v}$. In this work, we take $f_v$ as vulnerable functions and $f_p$ as non-vulnerable functions. We can get a tuple ($f_v$, $f_p$, $S_{del}$, $S_{add}$), where $S_{del}$, $S_{add}$ will be utilized for augmentation (See Section~\ref{sec:aug}). 

\begin{figure}[t]
    \centering
    \includegraphics[width=0.45\textwidth]{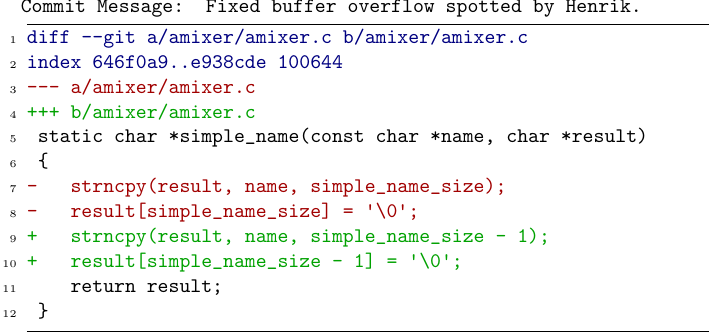}
    \caption{Patch for Buffer Overflow.}
    \label{fig:patch}
\end{figure}

An illustrative example of a security patch is shown in Figure~\ref{fig:patch}, we can get the changed statements $S_{del}$ and $S_{add}$ at line 7 to line 8 and line 9 to line 10, respectively. The vulnerable function $f_v$ is composed from line 5 to line 8 and line 11 to 12 in Figure~\ref{fig:patch}, and the patched function $f_p$ is composed from line 5 to line 6 and line 9 to line 12.

\vspace{2mm}





\subsection{Vulnerability-preserving Data Augmentation}\label{sec:aug}
As we collect different types of vulnerability data, it is difficult to ensure each type has a sufficient number for models to learn, hence we propose a data augmentation technique to scale up the collected data $D$ in Section~\ref{sec2b}. Furthermore, the newly generated data must retain the vulnerability characteristics of the original data, i.e., it needs to be vulnerability-preserving. If the vulnerability is lost or compromised, the generated data becomes ineffective for model training. Hence, we propose a novel vulnerability-preserving data augmentation method to generate new data from the original dataset $D$. It primarily involves two steps. The first step (Section~\ref{sec3.3.1}) is to slice all the statements related to the vulnerability. The second step (Section~\ref{sec3.3.2}) is to augment the original dataset $D$ by preserving the semantics of vulnerability-related statements and modifying the statements unrelated to the vulnerability.


\subsubsection{Slicing vulnerability-related statements}\label{sec3.3.1} 
Given a 4-tuple ($f_v$, $f_p$, $S_{del}$, $S_{add}$) from Section~\ref{sec3.2.2}, we slice the statements that are related to $S_{del}$ and $S_{add}$. To achieve this, we need to get the statement dependency relationship in a function. We utilize the program dependency graph (PDG) to obtain the data dependency and control dependency for each statement in $f_v$ and $f_p$. The generated PDGs are defined as $\mathrm{PDG}_{f_v}$ and $\mathrm{PDG}_{f_p}$. Based on PDG, we design an Algorithm~\ref{alg:slice} to slice the vulnerability-related statements. Specifically, for each statement $\mathrm{s}_{del}/\mathrm{s}_{add}$ in $S_{del}/S_{add}$, a forward slicing procedure is performed in PDG to get a list of related statements $\mathrm{S}_f$ in the function $f_{v}/f_{p}$. This step is to ensure find out the \textit{future} dependent statements from the current statement i.e., \textit{start from current statement}. 
Then based on the obtained $\mathrm{S}_f$, a backward slicing procedure is conducted to extract all relevant statements \textit{before} the $\mathrm{S}_f$ i.e., \textit{point to current statement}. 
Finally, we combine both directions for the added statements $\mathrm{S}_{add}$ and deleted statements $\mathrm{S}_{del}$ and obtain the vulnerability-related statements denoted as $\mathrm{S}_{related}$.

\vspace{2mm}
\begin{algorithm}[t]
\scriptsize
\caption{Vulnerability-related Slicing} 
\label{alg:slice}
\KwIn{($\mathrm{PDG}_{f_{v}}$, $\mathrm{PDG}_{f_{p}}$, $\mathrm{S}_{del}$, $\mathrm{S}_{add}$)} 
\KwOut{$\mathrm{S}_{related}$}
\SetKwFunction{FMain}{Slice}
\SetKwFunction{Fbw}{traverse}
\SetKwFunction{Fgr}{slice}
Initialize $\mathrm{S}_{related}$ = set()\\
\SetKwProg{Fn}{Function}{:}{}
\Fn{\FMain{$\mathrm{PDG}_{f_{v}}$, $\mathrm{PDG}_{f_{p}}$, $\mathrm{S}_{del}$, $\mathrm{S}_{add}$}}{
    \For{$\mathrm{s}_{del}$ in $\mathrm{S}_{del}$}
    {
        $\mathrm{S}_f$ = traverse ($\mathrm{s}_{del}$, $\mathrm{PDG}_{f_{v}}$, ``forward'')\\
        \For{s in $\mathrm{S}_f$}
        {
            $\mathrm{S}_b$ = traverse (s, $\mathrm{PDG}_{f_{v}}$, ``backward'')\\
            $\mathrm{S}_{related}$ = $\mathrm{S}_{related}$ $\cup$ $\mathrm{S}_b$\\
        }
    }
    \For{$\mathrm{s}_{add}$ in $\mathrm{S}_{add}$}
    {
        $\mathrm{S}_f$ = traverse ($\mathrm{s}_{add}$, $\mathrm{PDG}_{f_{p}}$, ``forward'')\\
        \For{s in $\mathrm{S}_f$}
        {
            $\mathrm{S}_b$ = traverse (s, $\mathrm{PDG}_{f_{p}}$, ``backward'')\\
            $\mathrm{S}_{related}$ = $\mathrm{S}_{related}$ $\cup$ $\mathrm{S}_b$\\
        }        
    }
}
\SetKwProg{Fn}{Def}{:}{}
\Fn{\Fbw{s, pdg, direction}}{
    results = set(), visited = set()\\
    q = Queue()\\
    q.push(s)\\
    \While{q is not empty}{
        u = q.pop()\\
        results.append(u)\\
        \If{direction == ``forward''}
        {
            $\mathrm{S}_{f}$ = the\ statements\ that\ start\ from\ u\ in\ $pdg$ \\
        }
        \ElseIf{direction == ``backward''}
        {
            $\mathrm{S}_{b}$ = the\ statements\ that\ point\ to\ u\ in\ $pdg$\\
        }
        \For{v in \{$\mathrm{S}_{f}$, $\mathrm{S}_{b}$\}}
        {
            \If{v $\notin$ $\mathrm{visited}$}
                {
                    visited.add(v)\\
                    q.push(v)\\
                }
        }
    }
    \KwRet results\\
}
\end{algorithm}

\subsubsection{Augmenting code by operators.}\label{sec3.3.2}

As all vulnerability-related statements i.e., $S_{related}$ are obtained, we can augment the data on the vulnerability-unrelated statements from the vulnerable function $f_v$ i.e., $\{s | s \in f_v \setminus S_{related}\}$ where $s$ is the vulnerability-unrelated statement and $\setminus$ is the set difference operation between $f_v$ and $S_{related}$. Preserving all vulnerability-related statements in the vulnerable function can retain its vulnerability and the proof is produced in Section~\ref{sec:proof}. We define five types of mutation operations for vulnerable data augmentation, as shown in Table~\ref{tb1-mutation}. 
Specifically, the operation \textit{rn} means to rename the used variable names with all the occurrences of these variables with other names. The operation \textit{ai} means to add a \textit{if} condition which is the logical true before the assignment statements. For example, suppose an assignment statement \textit{int a = b;} is vulnerability-unrelated, it can be transposed to \textit{if (True) then int a = b;} after performing the $ai$ operation. Operation \textit{del} will randomly delete the statements that are not related to vulnerability-related statements, while the \textit{add} will rename variable names in an assignment statement and add it back to the original function $f_v$, and the operation \textit{ro} means to reorder the consequent assignment statements in the original function. 
By different types of mutation operators, we can greatly increase the amount of original data and enrich its diversity.


\vspace{2mm}
\begin{table}[!t]
\centering
\caption{Mutation Operations for Data Augmentation.}
\scriptsize	
\begin{tabular}{ll}
\toprule 
\textbf{Type} & \textbf{Definition} \\ \midrule
\textit{add} &
\begin{tabular}[c]{@{}l@{}} Rename identifiers in a assignment statement and \\add it back to the function. \end{tabular}  \\ \midrule
\textit{ai} & 
\begin{tabular}[c]{@{}l@{}} Add a \textit{if} condition which is the logical truth before \\ assignment statements. \end{tabular}  \\ \midrule
\textit{rn} & Rename the identifiers.\\ \midrule
\textit{ro} & Reorder assignment statements.  \\ \midrule
\textit{del} & Delete statements that are not related to the vulnerability.
\\ \bottomrule 
\end{tabular}
\label{tb1-mutation}
\end{table}

\subsection{Edge-aware GGNN}
While GGNN has found extensive application in modeling source code~\cite{learning_to_represent, zhou2019devign, fernandes2018structured, liu2021retrievalaugmented, liu2023graphsearchnet}, it is noteworthy that the message passing is solely based on the node representations, i.e., $\boldsymbol h_{n_i}$, and the edge information is overlooked. We believe that the different types of edges in the Code Property Graph (CPG), such as "Flow to" and "Control" signify different semantics of programs, playing a crucial role in vulnerability detection. Building on this insight, we propose an edge-aware GGNN to leverage edge information effectively for vulnerability detection.

\subsubsection{Graph Initialization}

For both vulnerable and non-vulnerable functions, we utilize Joern~\cite{yamaguchi2014modeling} to obtain the Code Property Graph (CPG). In a formal representation, a raw function $c$ can be expressed as a multi-edged graph $g(\mathcal{V},\mathcal{E})$, where $\mathcal{V}$ is the set of nodes, and $(v, u) \in \mathcal{E}$ denotes the edge connecting node $v$ and node $u$. Each node possesses its node sequence, parsed by Joern from the original function. We tokenize the node sequence by spaces and punctuation. Additionally, for compound words (tokens constructed by concatenating multiple words according to camel or snake conventions), we split them into multiple tokens. We represent each token in the node sequence and each edge type connected with nodes using the learned embedding matrix $\boldsymbol E^{seqtoken}$ and $\boldsymbol E^{edgetype}$, respectively. Subsequently, the nodes and edges of the Code Property Graph (CPG) can be encoded as:
\begin{equation}
\small
\begin{gathered} 
\boldsymbol h_v = \mathrm{SUM}(\boldsymbol E^{seqtoken}_{v,1},...,\boldsymbol E^{seqtoken}_{v,l}) \\
\boldsymbol e_{v,u} =  \boldsymbol E^{edgetype}_{v,u}~~if~~(v, u) \in \mathcal{E}~~else~~\boldsymbol 0
\end{gathered}
\end{equation}
where $l$ denotes the number of tokens in the node $v$. Hence, given the code property graph $g(\mathcal{V},\mathcal{E})$, we have $\boldsymbol H \in \mathbb{R}^{m \times d} $, which denotes the initial node embedding matrix of the CPG, where $m$ is the total number of nodes in the CPG and $d$ is the dimension of the node embedding.  
\subsubsection{Edge-aware Message Passing}
For every node $v$ at each computation iteration $k$ in the graph, we employ an aggregation function to calculate the aggregated vector $\boldsymbol h_{\mathcal{N}_{(v)}}^k$. This is achieved by considering a set of neighboring node embeddings, as well as the connected edge type information computed from the previous hop. As the edge information is also taken into account in the message passing process, it is specifically referred to as edge-aware message passing.
\begin{equation}
\small
  \boldsymbol h_{\mathcal{N}_{(v)}}^k = \mathrm{SUM}(\{\mathrm{Relu}(\boldsymbol W [\boldsymbol h_u^{k-1}; \boldsymbol e_{v, u}]) | \forall u \in \mathcal{N}_{(v)} \})
\end{equation}
where $\mathcal{N}_{(v)}$ is a set of the neighboring nodes which are directly connected with $v$,  $\boldsymbol W \in \mathbb{R}^{(d + d') \times d} $ where $d$ and $d'$ are the dimension of the node and edge embedding, and Relu is the rectified linear unit~\cite{nair2010rectified}.
For each node $v$, $\boldsymbol h_v^0$ is the initial node embedding of $v$, i.e., $\boldsymbol h_v \in \boldsymbol {H}$.

A Gated Recurrent Unit (GRU)~\cite{cho2014learning} is used to update the node embeddings by incorporating the aggregation information.
 \begin{equation}
\small
 \boldsymbol h_v^k =\mathrm{GRU}(\boldsymbol h_v^{k-1}, \boldsymbol h_{\mathcal{N}_{(v)}}^k)
 \end{equation}
 After $n$ iterations of computation, we obtain the final node state $\boldsymbol h_v^n$ for node $v$. Subsequently, we apply max-pooling over all nodes ${\boldsymbol h_v^n | \forall v \in \mathcal{V} }$ to acquire the $d$-dimensional graph representation $\boldsymbol h^g$.
 \subsubsection{Classification Layer}
After the message passing, we can get the graph representation $\boldsymbol h^g$ and use it for prediction. Specifically, a liner projection with a sigmoid activation function is used to make the final prediction. 
 \begin{equation}
 \small
 \label{eq:logit}
  y' = \mathrm{Sigmoid}(\boldsymbol W' \boldsymbol h^g )
 \end{equation}
 where $y'$ is the logit produced by the sigmoid function and $\boldsymbol W' \in \mathbb{R}^{d \times 1} $ is the learned matrix.
\subsection{Training}
In \tool, for each type of vulnerability (refer to Table~\ref{keywords}), on the corresponding $D_{train_i}$, which is the training set containing vulnerable and non-vulnerable functions for the vulnerability type $i$, we train a set of binary classifiers $F = \{F_i$, $\forall i \in \mathrm{Vul}\}$, where $ \mathrm{Vul} = \mathrm{CWE}-\{404, 835, 120, 672, 362\}$ is the vulnerability type list to detect whether the function is vulnerable or not. The loss function for $F_i$ is binary cross entropy.
  \begin{equation}
 \begin{aligned}
 l(y', y) = -(\cdot y \cdot \mathrm{log}(y') + (1-y) * \mathrm{log}(1-y'))
 \end{aligned}
 \end{equation}
 where $y'$ is the logit (See Equation~\ref{eq:logit}) and $y \in \{0, 1\}$ is the label with 1 for vulnerable and 0 otherwise. Totally, we have five classifiers according to different vulnerability types.
\subsection{Testing}
Since \tool targets fine-grained vulnerability detection, we train each type of vulnerability as a single binary classifier $F_i$ and vote to give the final prediction for a test sample. Specifically, given a function $f_v$ (resp. $f_p$) from the test set, each type of classifier $F_i$ is employed for detection $y'_i=F_i(f_v)$ and the predicted label can be expressed as follows:
  \begin{equation}
\small
 \mathrm{Predicted\_label} =
    \begin{cases}
      1 &  y'_i > 0.5 \\
      0 &  y'_i \leq 0.5
    \end{cases}       
 \end{equation}
where 1 for vulnerable and 0 for non-vulnerable. The final prediction result is determined through a majority voting mechanism from all classifiers based on the majority rule.

\section{Evaluation Setup}\label{sec:evaluation}
\vspace{2mm}
\begin{table*}[!t]
\centering
\caption{The Statistics of the Collected DataSet.}
\scriptsize	
\begin{tabular}{c|cccc|ccc|c}
\toprule
\multirow{2}{*}{CWE} & \multirow{2}{*}{Commit} & \multirow{2}{*}{Function} & \multirow{2}{*}{Graph} & \multirow{2}{*}{Preprocess} & \multicolumn{3}{c|}{Raw dataset} & \multicolumn{1}{c}{Mutation} \\ \cmidrule{6-9} 
                     &                         &                           &                        &                             & train      & validation      & test     & rn/del/add/ai/ro    \\ \midrule
CWE-404              & 39,261                   & 78,522                     & 67,860                  & 41,816                       & 25,060      & 8,400       & 8,356     & 12,552               \\ 
CWE-835              & 16,584                   & 33,168                     & 29,904                  & 16,105                       & 9,638       & 3,263       & 3,204     & 4,839                  \\ 
CWE-120              & 10,877                   & 21,754                     & 19,800                  & 11,187                       & 6,710       & 2,250       & 2,227     & 3,370                 \\ 
CWE-672              & 7,906                    & 15,812                     & 15,006                  & 8,689                        & 5,197       & 1,741       & 1,751     & 2,604                 \\ 
CWE-362              & 17,897                   & 35,794                     & 32,652                  & 21,279                       & 12,768      & 4,247       & 4,264     & 6,390               \\ \midrule
Total                &92,525                  &  185,050               &165,222        &   99,076     & 59,373  &     19,901   &  19,802  &       29,755                                                                                        
\\ \bottomrule
\end{tabular}
\label{tb3}
\end{table*}

\begin{itemize}[leftmargin=*]
    \item \textbf{RQ1}: What is the performance of \tool compared with baselines in detecting vulnerable code?
    \item \textbf{RQ2}: Can each type of the defined mutation operations be beneficial to augment the training dataset to improve the detection accuracy?
    \item \textbf{RQ3}: What is the performance of our designed edge-aware GGNN compared with other GNN variants for vulnerability detection?
\end{itemize}

\subsection{Dataset Details}
The statistics for the five common vulnerability types on the collected dataset are presented in Table~\ref{tb3}. We first collect a total of 92,525 commits of the five CWE types, then extract vulnerable and patched functions from each commit as vulnerable and non-vulnerable functions. After that, we extract the code property graph (CPG) for each function and obtain 165,222 graphs in total, which is less than the number of the raw functions due to the compilation errors of some functions with Joern~\cite{yamaguchi2014}. We further conduct a data preprocessing to remove functions whose number of graph nodes is greater than 800, and finally obtain a raw data set with a total number of 99,076 functions. We divide the raw dataset into a train set, validation set, and test set at a ratio of 6:2:2. In the end, we perform five types of mutation operations (see Table~\ref{tb1-mutation}) to augment the vulnerable functions in the train set and generate the mutated functions for each type of mutation operations with a nearly equal amount of vulnerable functions. Note that, our dataset is more challenging than Devign~\cite{zhou2019devign}. In particular, the extracted non-vulnerable functions in Devign~\cite{zhou2019devign} come from non-vulnerable commits. However, \tool uses the fixed version of the code from the vulnerability-related commits as the non-vulnerable functions for the model to learn. As the non-vulnerable functions are highly similar to the vulnerable functions in this operation compared with Devign, hence it is a more difficult data set for DL-based approaches to learn vulnerability features to distinguish them.

\subsection{Baselines}
We evaluate \tool by comparing it against several well-known vulnerability detection approaches. 
\subsubsection{Static-analysis-based approaches}

\textbf{VUDDY}~\cite{kim2017vuddy}.
It follows a process to abstract the function and then generates fingerprints for each function by hashing the normalized code. A target function is identified as vulnerable if its fingerprints match those of vulnerable functions. 

\noindent\textbf{MVP}~\cite{xiao2020mvp}.
It is similar to VUDDY, which extracts vulnerability and patch signatures from a vulnerable function and its patched counterpart using a proposed program-slicing algorithm. Then It identifies a target function as vulnerable if it matches the vulnerability signature but does not match the patch signature. 

\subsubsection{Deep-learning-based approaches} 
\noindent\textbf{Vuldeepecker}~\cite{li2018vuldeepecker}.
It introduces a methodology involving extracting semantically connected statements associated with an argument of a library/API function call, forming code gadgets. Following program normalization, which standardizes user-defined variable names and function names, a bidirectional LSTM neural network is employed to determine the function's vulnerability.

\noindent\textbf{Multi-Head Attention}~\cite{vaswani2017attention}.
It has been widely used for modelling sequences. In particular, we leveraged the documentation from harvardnlp~\cite{opennmt} to construct a multi-head attention layer, setting the number of heads to 4 and the maximum sequence length to 150 for comparative analysis.

\noindent\textbf{Devign}~\cite{zhou2019devign}.
It is a typical work in vulnerability detection utilizing graph neural networks. Specifically, it combines varied semantics of a function into a unified graph structure to glean program semantics. Additionally, it employs a convolution module to capture features related to vulnerabilities.

\noindent\textbf{CodeBERT}~\cite{feng2020codebert}.
It is a pre-trained model rooted in the Transformer architecture for code modeling. Leveraging millions of code data, it undergoes pre-training and subsequent fine-tuning for downstream code-related tasks. We have reproduced its implementation using the default settings provided in the official code on our dataset.



\subsection{Experimental Settings}
We utilized the common words where the frequency $\geq$ 3 from the training set, amounting to 90,000, to create the vocabulary set. The dimensions of word and edge embeddings were set to 128 respectively. A dropout of 0.3 was implemented after the word embedding layer. The hop value was set to 4 for CWE-404 and CWE-672, 1 for CWE-835, 5 for CWE-120, and 2 for CWE-362 to achieve optimal performance. We employed the Adam optimizer with an initial learning rate of 0.001 and a batch size of 64 for training. All experiments were conducted on the DGX server with three Nvidia Graphics Tesla V100.

\vspace{2mm}
\begin{table*}[!t]
\caption{The experimental results of different approaches for vulnerability detection.}
\label{tbl-baselines}
\centering
\scriptsize	
\begin{tabular}{c|ccc|ccc|ccc|ccc|ccc}
\toprule
\multirow{2}{*}{Approach}           & \multicolumn{3}{c|}{CWE-404}                                                       & \multicolumn{3}{c|}{CWE-835}                                                       & \multicolumn{3}{c|}{CWE-120}                                                       & \multicolumn{3}{c|}{CWE-672}                                                       & \multicolumn{3}{c}{CWE-362}                                                                                                                  \\ \cmidrule{2-16} 
& P     & R     & F1     & P      & R     & F1    & P     & R     & F1   & P   & R    & F1  & P & R    & F1           \\ \midrule
VUDDY  & \textbf{73.17} & 0.72& 1.42  & \textbf{70.00} & 0.43 & 0.86  & \textbf{79.41}  & 2.42 & 4.70 & \textbf{84.62} & 1.25 & 2.47 & \textbf{60.00} & 0.28 & 0.56   \\ 
MVP & 49.89  & 45.30 & 47.48  & 50.68   & 46.15  & 48.31 & 50.19 & 48.21 & 49.18  & 50.13  & 43.17  & 46.39 & 49.87  & 45.60  & 47.64  \\ \midrule
Vuldeepecker & 67.26 & 38.19 & 48.71  & 50.70    & 58.52  & 54.33    & 50.55 & 49.51 & 50.02 & 51.39    & 56.95 & 54.02  & 51.57 & 43.64 & 47.28\\ 
Attention   & 56.82 & 57.66 & 57.24 & 50.94 & 65.22 & 57.21 & 55.1 & 57.91 & 56.47 & 52.79 & 52.73 & 52.76 & 51.38 & 53.75 & 52.54  \\ 
Devign  & 65.86 & 40.98 & 50.52 & 50.79 & 61.50 & 55.64  & 55.30 & 63.90 & 59.29 & 51.56& 43.17  & 46.99 & 51.16 & 44.01 & 47.32  \\ 
CodeBERT & 64.59 & 54.46& 59.10 & 51.72 & 64.29 & 57.32 & 60.40 & 35.03  & 44.34 & 52.69 & 63.55 & 57.61 & 52.69& 60.32 & 56.24  \\ \midrule
GGNN & 64.96 & 46.01 & 53.87 & 51.64 & 59.70 & 55.38 & 55.71 & 60.14 & 57.84 & 51.67 & 60.02 & 55.53 & 53.49 & 55.01 & 54.24  \\ 
GCN  & 65.17 & 40.07 & 49.63 & 50.82 &76.88 & 61.19 & 55.72 &64.88 & 59.95 & 52.99 &60.59 & 56.54 & 52.76 &54.68 & 53.71 \\ \midrule
$\tool_{none}$  & 64.45 & 46.95 & 54.32 & 51.07 & 66.58 & 57.80 & 56.52 & 70.15 & 62.60 & 50.54 & 69.82 & 58.63 & 52.10 & 72.66 & 60.69 \\ 
\tool & 53.91 & \textbf{81.15} & \textbf{64.78} & 50.51 & \textbf{94.92} & \textbf{65.93} & 52.93 & \textbf{86.33} & \textbf{65.63} & 50.53 & 92.26 & 65.30 & 50.47 & \textbf{90.87} & \textbf{64.89} \\ \hdashline
$\tool_{add}$ & 61.95 & 49.74 & 55.18 & 50.60 & 73.53 & 59.94 & 55.25 & 73.37 & 63.03 & 50.90 & 77.45 & 61.43 & 50.97 & 85.89 & 63.97  \\ 
$\tool_{ai}$  & 62.58 & 54.51 & 58.27 & 50.77 & 71.11 & 59.25 & 56.71 & 64.61 & 62.80 & 50.72 & 72.21 & 59.59 & 51.47 & 70.94 & 61.66 \\ 
$\tool_{del}$ & 60.80 & 58.40 & 59.57 & 51.47 & 62.86 & 56.60 & 56.28 & 66.49 & 60.96 & 52.48 & 55.47 & 53.93 & 51.57 & 67.26 & 58.38  \\ 
$\tool_{rn}$  & 53.41 & 78.31 & 63.51 & 50.49 & 91.94 & 65.19 & 53.20 & 83.82 & 65.09 & 50.45 & \textbf{96.47} & \textbf{66.25} & 51.03 & 87.80 & 64.54  \\ 
$\tool_{ro}$  & 60.38 & 58.71 & 59.54 & 51.41 & 76.81 & 61.60 & 55.32 & 67.83 & 60.94 & 50.18 & 77.45 & 60.90 & 52.05 & 63.30  & 57.12 \\  \bottomrule
\end{tabular}
\end{table*}

\section{Experimental Results}\label{sec:results}
\subsection{RQ1: Compared with Baselines}
The experimental results are presented in Table~\ref{tbl-baselines}. The first row presents the results for the static-analysis-based approaches, the second row is the results for DL-based approaches. The results for \tool without/with the data augmentation are provided in the row of \tool$_{none}$ and \tool respectively. 

When comparing the results of \tool with static-analysis-based approaches, it is obvious that VUDDY achieves much higher precision scores. For instance, in the case of the vulnerability CWE-672 (Operation After Free), VUDDY achieves a precision score of 84.62 which is much higher than the score of \tool 50.53. The higher precision score indicates that VUDDY has fewer false positive samples, which is reasonable as VUDDY relies on expert-crafted features to detect code vulnerabilities. These hand-crafted vulnerability features are highly reliable by security experts. Therefore, if the samples being detected exhibit similar features, there is a high probability that they are vulnerable code. However, we can also find that VUDDY has a lower recall than \tool i.e., 1.25 vs 92.26. It indicates that VUDDY has more false negative samples as these hand-crafted vulnerability features can only cover a limited number of vulnerability types, which leads to missing a substantial number of vulnerabilities compared to \tool. In addition, we find that although MVP has lower precision scores than VUDDY, its recall scores are better than VUDDY, which indicates that MVP covers more vulnerability types but the detection precision is lower than VUDDY. Compared with these static-analysis-based approaches, \tool is able to achieve a much higher recall, which leads to a higher F1-score.

When comparing the results of \tool with the DL-based approaches, we can find that the pre-trained model CodeBERT performs better than other baselines in terms of F1. We speculate that the main reason is that CodeBERT uses extensive code-related data for pre-training and the model architecture is more powerful than the other baselines. Hence, CodeBERT has a stronger learning capability. However, \tool outperforms it in terms of recall and F1. Even without data augmentation i.e., \tool$_{none}$ in Table~\ref{tbl-baselines}, we can find that it still has better performance than CodeBERT in terms of F1 for vulnerability types CWE-{835, 120, 672, 362), which indicates the effectiveness of our proposed approach. 

\begin{tcolorbox}[breakable,width=\linewidth,boxrule=0pt,top=1pt, bottom=1pt, left=1pt,right=1pt, colback=gray!20,colframe=gray!20]
\textbf{Answer to RQ1:} Although some static-analysis-based approaches have higher precision scores than \tool, they have extensive false negative samples. Overall, in terms of Recall and F1, \tool outperforms current baselines including static-analysis-based approaches and DL-based approaches by a significant margin.
\end{tcolorbox}

\subsection{RQ2: Effectiveness of Mutation Operations}\label{sec:rq2}
In our work, we introduce five types of mutation operations to augment the data. We assess the effectiveness of each operation individually by conducting experiments using only one type of mutation operation at a time, while maintaining the hyper-parameters consistent with the original model. The results of these experiments are outlined in the final row of Table~\ref{tbl-baselines}, where $\tool_{{*}}$ denotes the specific type of mutation operation being evaluated. The combined results of all five mutation operations are presented in the row labeled $\tool$.

Through the analysis of the experimental results, it is evident that incorporating five types of mutation operations can significantly improve recall and F1 scores. For instance, in the case of vulnerability type CWE-404 (indicating memory leaks), \tool improves recall and F1 from 46.95/54.32 to 81.15/64.78, respectively. Notably, the \textit{rn} operation stands out as the most effective in enhancing F1 across different vulnerability types.  Even for the vulnerability CWE-672, when fusing other mutation types of data to \textit{rn}, F1 has a decrease. We conjuncture that \textit{rn} operation appears to be efficient in improving the diversity of the training set compared to other mutation operations, making the model more robust and powerful.

Additionally, different types of mutations exhibit inconsistent performance compared to the original model without mutations ($\tool_{none}$). For example, the mutation operations \textit{add}, \textit{ai}, and \textit{rn} improve F1 for all types of vulnerabilities compared to $\tool_{none}$. However, the \textit{del} operation has a negative impact except for CWE-404 (memory leak), while \textit{ro} improves F1 for vulnerability types CWE-\{404, 835, 672\} but has a negative impact for CWE-\{120, 362\} compared to $\tool_{none}$. This may be attributed to the fact that the \textit{del} and \textit{ro} operations introduce some other types of vulnerabilities in the mutated functions, adding noise to the dataset and making it challenging for the model to make correct decisions. For more details about the reason that \textit{del} and \textit{ro} can introduce new vulnerabilities, please refer to Section~\ref{sec:proof}. Despite the negative impact of \textit{del} and \textit{ro} on specific vulnerability types, combining them with other mutation operations in \tool yields the best overall performance.

\begin{tcolorbox}[breakable,width=\linewidth,boxrule=0pt,top=1pt, bottom=1pt, left=1pt,right=1pt, colback=gray!20,colframe=gray!20]
\textbf{Answer to RQ2:} A thorough analysis of the performance of different mutation operations leads us to the conclusion that vulnerability-preserving data augmentation is effective for further enhancement.
\end{tcolorbox}

\subsection{RQ3: Effectiveness of Edge-aware GGNN}
\tool proposes the edge-aware GGNN which extends the current GGNN by encoding the edge-type information and using it in the message passing process to learn the vulnerability-related features. To illustrate the effectiveness of the proposed model, we also compared it with some GNN variants such as Gated Graph Network (GGNN) and Graph Convolution Network (GCN). The experimental results are shown in Table~\ref{tbl-baselines}.

To make a fair comparison, we only compare the results of different GNN variants with {$\tool_{none}$}. We find that compared with GGNN, supplementing the edge type information is beneficial for the model to detect vulnerabilities. It is reasonable as different types of vulnerabilities are involved in different aspects of the program i.e., in the different types of code property graph. For example, the vulnerability of CWE-120 (Buffer Overflow) is about the definition and usage of a variable, which can be captured in the program dependency graph via the edges of ``Define/Use''. However, the vulnerability of CWE-672 (Operation After Free) is related to the execution order of the statements and it is reflected in the control flow graph through the edges of ``Flow to''. Thus, we embed the edge type information explicitly and use it for learning will decrease the difficulty of the model in identifying the vulnerability and improve the model's capability. Furthermore, we also compare with GCN, which uses multiple graph convolutional layers in updating node representations, $\tool_{none}$ still has a better performance excluding CWE-835 in terms of F1 score. 

\begin{tcolorbox}[breakable,width=\linewidth,boxrule=0pt,top=1pt, bottom=1pt, left=1pt,right=1pt, colback=gray!20,colframe=gray!20]
\textbf{Answer to RQ3:} Edge-aware GGNN, which encodes the edge type information explicitly for learning can increase the model capability and reduce the difficulty for the model to detect vulnerabilities, thus it can produce a better performance compared with some other GNN variants i.e., GGNN, GCN.
\end{tcolorbox}

\section{Discussion}\label{sec:discussion}
\subsection{Dataset Reliability}\label{sec:data_reliability}
To ensure the reliability of our dataset, we established a systematic data collection pipeline. Initially, we crawled a substantial number of commits from 1614 C-language open-source projects on GitHub. Subsequently, we filtered out commits that only modified a single function and whose messages were matched by one vulnerability type (refer to Section~\ref{sec2b}). By employing this method, the functions extracted from the commits are more likely to be vulnerabilities with the correct vulnerability type. To validate the collected dataset, a team of professional security researchers randomly selected 400 commits from each of the 5 vulnerability types, totaling 2000 commits, and conducted a two-round cross-verification on the data labeling. The results revealed that 97.3\% of commits were correctly classified (CWE-404 was 97.50\%, CWE-835 was 96.50\%, CWE-120 was 97.75\%, CWE-672 was 98.50\%, and CWE-362 was 96.25\%, respectively). The high precision of our dataset indicates that vulnerability-related commits can be effectively identified and correctly classified using keywords from an unlabeled dataset.

\subsection{Proof of Vulnerability-preserving}\label{sec:proof}
Here we discuss the proof of vulnerability-preserving augmentation, which can preserve the vulnerability of a function after being modified. First, we have the following notation $S_{vul}$ denotes the statements that trigger the vulnerability in a function, and $S_{vul_{dep}}$ denotes the dependent statements of the statements $S_{vul}$ in a function. We have the assumption:

\noindent \textbf{Assumption}: If all $S_{vul}$ and $S_{vul_{dep}}$ are retained after the mutation operations, then the vulnerability still exists.

The assumption is correct since the vulnerabilities are composed of $S_{vul}$ and $S_{vul_{dep}}$. Hence, as long as we prove that our proposed method retains all $S_{vul}$ and $S_{vul_{dep}}$, it is vulnerability-preserving.


We understand that commits in GitHub primarily serve to record code changes for different versions, and the commits extracted through our Data Collection process (Section~\ref{sec2b}) are identified as vulnerability-related through keyword filtering. The cross-validation results presented in Section~\ref{sec:data_reliability} confirm the reliability of these commits concerning vulnerabilities. Consequently, the changed statements $S_{del}$ and $S_{add}$ in the vulnerability commit are unequivocally associated with the vulnerability, i.e., $\exists S \subseteq {S_{del} \cup S_{add}} \Rightarrow S \subseteq { S_{vul} \cup S_{vul_{dep}}}$. This suggests that if we apply the vulnerability-related slicing algorithm~\ref{alg:slice} by iteratively searching for related statements via PDG to each statement in $S_{del}$ and $S_{add}$ separately, then all $S_{vul} \cup S_{vul_{dep}}$ can be encompassed, i.e.,  

\begin{equation}
\small
  \{S_{vul} \cup S_{vul_{dep}}\} \subseteq S_{related} 
\end{equation}



Once we obtain the vulnerability-related statements $S_{related}$, we can proceed with data augmentation by considering the set difference between the raw vulnerable function $f_v$ and $S_{related}$. This augmentation process ensures the preservation of vulnerability in the augmented functions.

However, it is important to acknowledge that the defined mutation operations cannot guarantee the absence of new vulnerabilities. Consider the scenario where a statement like "usbDevs = NULL;" is unrelated to vulnerabilities and is used to set the pointer ``usbDevs" to NULL to mitigate the vulnerability of ``use after free". If we delete the statement ``usbDevs = NULL;" using the mutation operation $del$, it could introduce the vulnerability CWE-672 (Operation After Free). Similarly, in the case of a shared variable among multiple threads, if its assignment statement is placed before the lock operation, using the mutated operation $ro$ might result in a vulnerability CWE-362 (race condition).
Fortunately, we observe that the proposed operations $rn$, $add$, and $ai$ do not introduce new vulnerabilities. Only two operations, $del$, and $ro$, have the potential to generate new vulnerabilities. Moreover, as indicated in Section~\ref{sec:rq2}, the combination of all five mutation operations in \tool yields the highest performance. In summary, ensuring the absence of new vulnerabilities when modifying the semantics of the original function is a significant challenge, and addressing this remains in future.

\vspace{-1mm}
\subsection{Threats to Validity}
The first potential threat concerns the limited scope of the collected dataset, which includes only five types of vulnerabilities in the C programming language. While this might not cover all possible vulnerabilities, we assert that the selected vulnerability types are common and capable of causing significant harm to software systems. Moreover, the proposed model is not restricted to this specific dataset and can be extended to detect other types of vulnerabilities. Another potential threat is the exclusion of functions with node sizes exceeding 800 in the graph. This practice is consistent with Devign~\cite{zhou2019devign} due to the GPU memory requirements of GNNs. Truncating the graph size is a pragmatic choice to facilitate experiments, as GNNs necessitate substantial GPU memory.

\section{Related Work}\label{sec:related}
\noindent \textbf{Vulnerability Detection}
Static analysis plays a crucial role in identifying flaws and errors in a codebase to prevent the introduction of vulnerabilities and security bugs. Commercial tools like flawfinder~\cite{flawfinder} and CPPCheck~\cite{cppcheck} provide extensive static code analysis, helping discover early bugs and vulnerabilities. 
VUDDY utilizes a clone-based approach by matching the signature of vulnerable functions with target program signatures. Another approach by Yang et al.~\cite{xiao2020mvp} employs novel slicing techniques while incorporating vulnerability signatures. However, these analyses often require significant context and the extensive efforts of experts. With the increasing popularity of deep learning approaches, various works, such as VulDeePecker~\cite{li2018vuldeepecker}, VulSniper~\cite{duan2019vulsniper}, and Devign~\cite{zhou2019devign}, have utilized deep learning techniques to predict and detect vulnerabilities. Saikat et al.~\citep{chakraborty2021deep} systematically study the existing deep-learning-based vulnerability detection approaches. 
Besides the function-level vulnerability detection, LineVD~\cite{10.1145/3524842.3527949} leverages graph neural networks to locate the buggy statements. LineVul~\cite{10.1145/3524842.3528452} employs the transformer to identify the line vulnerability. 
These methods typically use an intermediate representation of programs, such as tokens or graphs, to facilitate the learning of meaningful program representations. In contrast, our work introduces a vulnerability-preserving data augmentation process to enhance vulnerability data for training purposes, which offers a different perspective.

\noindent \textbf{Data Augmentation}
It is a valuable technique for expanding datasets during model training in various domains such as computer vision (CV), natural language processing (NLP), and automated speech recognition (ASR). This technique is particularly useful in domains with limited datasets to prevent overfitting, as seen in medical image analysis~\cite{shin2018medical}. Moreover, data augmentation has been employed to enhance model performance. Studies~\cite{perez2017effectiveness, fawzi2016adaptive} introduce new methods for data augmentation, aiming to achieve better performance in image classification tasks. In code representation, Jain et al.~\cite{jain2020contrastive} and Liu et al.~\cite{liu2023contrabert} explore data augmentation in pre-training. Zhuo et al.~\cite{zhuo2023data} and Dong et al.~\cite{dong2023boosting} conduct the literature review of source code augmentation. The current strategies used are semantic-preserving at the functional granularity, which may not preserve vulnerability when generating mutations for code vulnerabilities. In contrast, our work introduces vulnerability-preserving data augmentation. Our approach augments the original dataset by preserving the semantics of vulnerability-related statements while modifying statements unrelated to vulnerabilities. It ensures that the generated variants retain vulnerability information, distinguishing it from other semantic-preserving strategies.

\section{Conclusion}\label{sec:conclusion}

In this paper, we introduce a fine-grained vulnerability detector namely \tool, which employs multiple classifiers to learn characteristics of various vulnerability types for source code vulnerability detection. To address the scarcity of data for some vulnerability types, we propose a novel vulnerability-preserving data augmentation technique for augmentation. Furthermore, we extend GGNN to an edge-aware variant to capture edge-type information. Extensive experiments have confirmed the effectiveness of \tool.

\section{Acknowledgment}
 This research is supported by the National Research Foundation, Singapore, and the Cyber Security Agency under its National Cybersecurity R\&D Programme (NCRP25-P04-TAICeN), the National Research Foundation, Singapore, and DSO National Laboratories under the AI Singapore Programme (AISG Award No: AISG2-GC-2023-008), and NRF Investigatorship NRF-NRFI06-2020-0001. Any opinions, findings and conclusions or recommendations expressed in this material are those of the author(s) and do not reflect the views of National Research Foundation, Singapore and Cyber Security Agency of Singapore.

\onecolumn
\begin{multicols}{2}
   \bibliographystyle{ACM-Reference-Format}
\bibliography{sample-base}
\end{multicols}





\end{document}